\documentclass[prl,twocolumn,english,superscriptaddress,floatfix]{revtex4}
\usepackage{graphicx}
\usepackage{bm, amsmath, amssymb}
\usepackage{color}
 \usepackage{soul}
\usepackage{pdfpages}
\usepackage[T1]{fontenc}
\usepackage[latin9]{inputenc}
\usepackage{amstext}
\usepackage{bbold}
\usepackage{esint}

\newcommand{\add}[1]{\textcolor{black}{#1}}
\newcommand{\addk}[1]{\textcolor{black}{#1}}
\def\be{\begin{equation}}
\def\ee{\end{equation}}
\def\bea{\begin{eqnarray}}
\def\eea{\end{eqnarray}}

\usepackage{babel}

\begin{document}

\title{\add{Quantum smoothing for classical mixtures}}
%\title{Quantum state--measurement signal correlations in continuous weak measurement}

%
%\author{N. Foroozani}
%\affiliation{Department of Physics, Washington University, St.\ Louis, Missouri 63130}

%%\affiliation{Quantum Nanoelectronics Laboratory, Department of Physics, University of California, Berkeley CA 94720}
\author{D. Tan}
\affiliation{Department of Physics, Washington University, St.\ Louis, Missouri 63130}
\author{M. Naghiloo}
\affiliation{Department of Physics, Washington University, St.\ Louis, Missouri 63130}
%\affiliation{Quantum Nanoelectronics Laboratory, Department of Physics, University of California, Berkeley CA 94720}
\author{K. M\o lmer}
\affiliation{Department of Physics and Astronomy, Aarhus University, Ny Munkegade 120, DK-8000 Aarhus C, Denmark}
\author{K. W. Murch}
\affiliation{Department of Physics, Washington University, St.\ Louis, Missouri 63130}
\affiliation{Institute for Materials Science and Engineering, St.\ Louis, Missouri 63130}
\date{\today}

%\author{Murch Group}
%\affiliation{Department of Physics, Washington University, St.\ Louis, Missouri 63130}
%\author{K. M\o lmer}
%\affiliation{Department of Physics and Astronomy, Aarhus University, Ny Munkegade 120, DK-8000 Aarhus C, Denmark}
%\date{\today}

\begin{abstract}
Wave functions and density matrices represent our knowledge about a quantum system and give probabilities for the outcomes of measurements. If the combined dynamics and measurements on a system lead to a density matrix $\rho(t)$ with only diagonal elements in a given basis $\{|n\rangle\}$, it may be treated as a classical mixture, i.e., a system which randomly occupies the basis states $|n\rangle$ with probabilities $\rho_{nn}(t)$. Equivalent to so-called smoothing in classical probability theory, subsequent probing of the occupation of the states $|n\rangle$ may improve our ability to retrodict what was the outcome of a projective state measurement at time $t$. Here, we show with experiments on a superconducting qubit that the smoothed probabilities do not, in the same way as the diagonal elements of $\rho(t)$, permit a classical mixture interpretation of the state of the system at the past time $t$.   \end{abstract}

\maketitle

%\section{Introduction}

The quantum mechanical wavefunction, $\psi(x)$, yields the probability for detection of a particle at location $x$, but most textbooks carefully emphasize that this does not imply that, prior to detection, the particle \emph{was} at the location $x$ with that probability. In contrast, a density matrix $\rho$ is often attributed a mixed interpretation as a classical random mixture of quantum states, i.e., the system is said to populate one out of several candidate states. A density matrix $\rho$  which is diagonal in a particular basis $|n\rangle$, indeed, leads to the same predictions about \add{the outcomes of} projective measurements in  that basis, $P(n)=\rho_{nn}$, as if states had been \add{assigned to the system with these probabilities.} Moreover, for any general measurement, described by a positive operator valued measure (POVM) \cite{Nielsenbook} with operators $\Omega_m$ that fulfill $\sum_m  \Omega_m^\dagger \Omega_m = I$  (the identity operator), the outcome probabilities $P(m) =$Tr($\Omega_m \rho\add{(t)} \Omega_m^\dagger$) equal the weighted mean of the probabilities over \add{a classical mixture of} states $|n\rangle$,
\begin {eqnarray} \label{eq:weighted}
P^{cm}(m) = \sum_n P(n) \textrm{Tr}\bigl(\Omega_m \rho_n \Omega_m^\dagger\bigr),
\end{eqnarray}
where $\rho_n=|n\rangle \langle n|$.

\add{When an experiment where data is collected over time has been completed, it is possible to examine the complete measurement record and use data obtained both before and after any time $t$ to yield information about the state of the system at $t$. In an analysis of classical stochastic processes we thus treat our (usual) knowledge about the system conditioned on earlier measurements as \emph{prior} probabilities which we update according to the later part of the data record by application of Bayes' rule \cite{numrec,speech}. The probability to obtain a given measurement data sequence between $t$ and the final probing time $T$ is conditioned on the state of the system at time $t$ and can be found by solving a recursive set of equations backwards from $T$ to $t$. The so-called forward-backward analysis \cite{numrec,speech} consists in determining the separate sets of prior and conditional probabilities and multiplying them according to Bayes' rule.}

\add{The quantum theory of measurements allows a similar analysis  of quantum processes, where the density matrix  $\rho(t)$, which depends on the evolution dynamics and measurements performed prior to time $t$, is supplemented by a positive, Hermitian matrix, denoted $E(t)$, which is calculated by a backward stochastic propagation equation from the final time $T$ until $t$ \cite{Gamm13}. The same way that $\rho(t)$ predicts the outcome probabilities for any hypothetical measurement, the pair of matrices $(\rho(t), E(t))$ exhaust our ability at time $T$ to assign outcome probabilities to any such measurement performed at the earlier time $t$ \cite{Gamm13},}
\begin{equation} \label{eq:pqsprob}
P_P(m) = \frac{\textrm{Tr}(\Omega_m \rho(t) \Omega_m^\dagger E(t))}{\sum_{m'} \textrm{Tr}(\Omega_{m'} \rho(t) \Omega_{m'}^\dagger E(t))}.
\end{equation}
\add{The subscript $P$ for \emph{Past} in Eq.\eqref{eq:pqsprob} recalls that we are (at time $T$ or later) retrodicting the probability for the outcome of a measurement at the past time $t$. The name \emph{quantum smoothing} has been proposed for the retrodiction of properties of the quantum systems \cite{Tsan09,Tsang09,Tsang11,Guev15,Armen09}, derived from the similar term smoothing used for classical stochastic processes. Indeed, the inferred probabilities tend to fluctuate less due to the accumulation of more relevant information and the correction, in hindsight, of mistaking statistical  signal fluctuations with actual transitions \cite{GammBonn}. Like the conventional quantum state $\rho(t)$, the pair $(\rho(t),\ E(t))$ of matrices is, notably, independent of the hypothetical measurement carried out at $t$, and following \cite{Gamm13} we shall denote it as the past quantum state (PQS).}

When applied to measurements on quantum systems, the PQS expression Eq.\eqref{eq:pqsprob} reveals unique features such as anomalous weak values \cite{Ahar88} arising from the pre- and postselection process \cite{Hatr13,murc13traj,Groe13,Lang14} and quantum coherence \cite{dres15,tan15}.  Smoothed predictions for the outcomes of measurements on quantum systems have been tested in a variety of experimental systems \cite{tan15,Ryba15,whea10} and they have been used in the interpretation of temporal signal correlation functions \cite{camp13, chan15, xu15, Foro16}.

\add{In this Letter, we study the particular case where the dynamics and the probing of the quantum system restrict the density matrix $\rho(t)$ and the matrix $E(t)$ to be diagonal in a definite basis $\{|n\rangle\}$.
In that case, Eq.\eqref{eq:pqsprob} yields the probability that a measurement at time $t$ found the system in state $|n\rangle$ ($\Omega_n = |n\rangle\langle n|$),}
\begin{equation} \label{eq:qfb}
P_P(n)=\frac{\rho_{nn}\add{(t)}E_{nn}\add{(t)}}{\sum_{n'}\rho_{n'n'}\add{(t)}E_{n'n'}\add{(t)}}
 \end{equation}
\add{which is, in  turn, completely equivalent to the expression obtained in the classical forward-backward analysis \cite{numrec,speech}.  In analogy with the interpretation of a diagonal density matrix $\rho(t)$, one might therefore expect that smoothed probabilities $P_P(n)$ would also permit a classical mixture interpretation as if the system \emph{did} occupy the quantum states $|n\rangle$ with probabilities $P_P(n)$ at time $t$.
But, the prediction based on \add{such a classical mixture interpretation of the state defined by the pair of diagonal matrices $\rho(t)$ and $E(t)$ }, cf. \eqref{eq:weighted},}
\begin{equation} \label{eq:cmprob}
P_P^{cm}(m)= \sum_{\add{n}} P_P(n) \textrm{Tr}\bigl(\Omega_m|n\rangle \langle n|\Omega_m^\dagger\bigr),
\end{equation}
\add{generally disagrees with Eq.\eqref{eq:pqsprob} for operators $\Omega_m$ which are not diagonal in the same basis as $\rho(t)$ and $E(t)$. The past quantum state or quantum smoothing theory does not merely replace the diagonal elements of $\rho(t)$ by another "more precise" set of probabilities, and no classical mixture interpretation can quantitatively account for both the measurements that are diagonal and not diagonal in the eigenbasis $\{|n\rangle\}$ of $\rho(t)$.}

\add{At this stage the reader may observe that actual test measurements will cause back-action on the quantum system and will, for some $\Omega_m$, populate states which are not diagonal in the eigenbasis $\{|n\rangle\}$ of $\rho(t)$. Our central question, however, is independent of specific test measurements and their back-action: It asks if our formally diagonal description of the system, valid as long as we are ignorant of the outcome of such actual measurements, is equivalent to a classical mixture.}

So far, we \addk{have} merely observed an inconsistency between different theoretical predictions for experiments.
We shall now present experiments on a superconducting qubit, where projective test measurements in bases different from the density matrix eigenbasis will illustrate and confirm Eq. \eqref{eq:pqsprob} while rejecting the classical mixture \add{interpretation} leading to Eq. \eqref{eq:cmprob}.

% \section{Experiment}

\begin{figure}\begin{center}
\includegraphics[angle = 0, width =.48\textwidth]{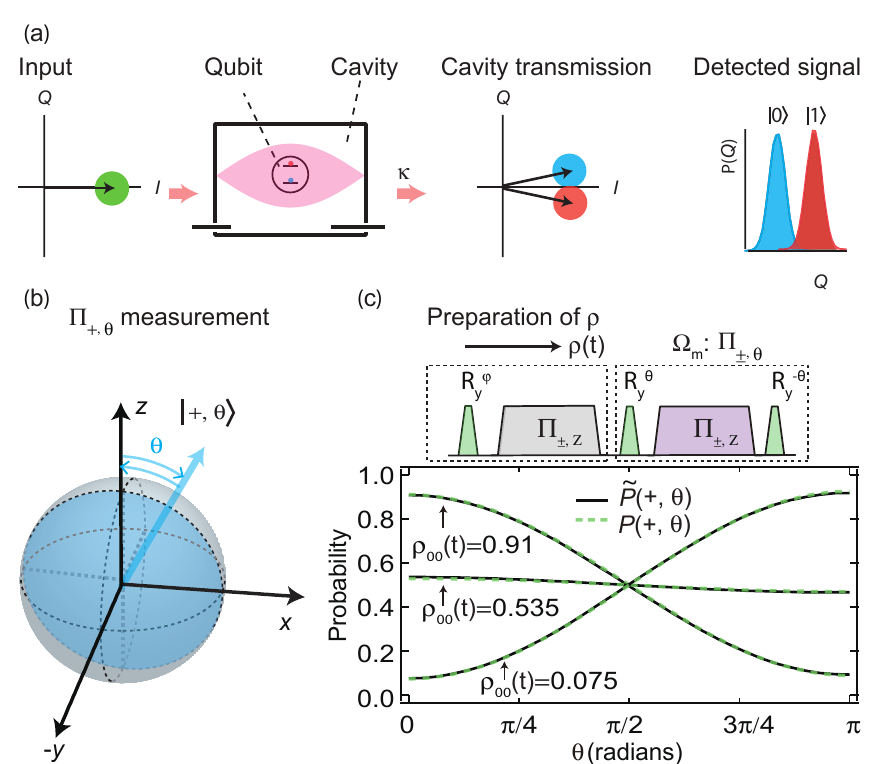}
\end{center}
\caption{ \label{fig1}  \add{(a)} The dispersive interaction between a superconducting qubit and a  cavity results in a qubit-state-dependent phase shift on a weak coherent drive on the cavity.  Sufficient drive strength and narrow integration bandwidth result in disjoint measurement distributions for one of the field quadratures, allowing single shot, quantum non-demolition measurements of the qubit in the energy eigenbasis.   (b) By combining projective measurements in the energy basis (along $z$) with rotations about the $y$ axis of the qubit state, projective measurements along an axis that forms an angle $\theta$ with the $z$ axis can be realized. (c)  Different initial states $\rho$ that are diagonal in the energy basis are prepared by performing an initial rotation and projective measurement. The results of the projective measurement are ignored.  We verify that the probability of a positive projective measurement outcome $\tilde{P}(+,\theta)$ is in agreement with the predictions of the initial density matrix  $P_\rho(+,\theta)$ for three different initial mixed states characterized by ($\rho_{00}\addk{(t)} =0.91,\ 0.535,\ 0.075$). \add{Over $5\times 10^4$ experimental repetitions are used for each measured  $\tilde{P}(+,\theta)$ leading to a statistical uncertainty of order $4\times10^{-3}$.}} \label{fig:rho}
\end{figure}

Our experiment, depicted in Figure \ref{fig:rho}a, consists of a superconducting transmon circuit that is dispersively coupled to \add{a 3D} aluminum  cavity \cite{EXP}.  \add{The anharmonicity of the transmon allows us to restrict the dynamics to the two lowest levels of the transmon \addk{realizing} a pseudo-spin half system described by a $2\times 2$ density matrix $\rho$.}  The dispersive interaction between the qubit and cavity is given by an interaction Hamiltonian $H_\mathrm{int.} = -\hbar \chi \sigma_z a^\dagger a$, \add{where $ \sigma_z$ (and $\sigma_x,\ \sigma_y$) are Pauli operators,} $a^\dagger (a)$ are the creation (annihilation) operators for a photon in the cavity mode and $\chi$ \add{is the dispersive coupling}. This interaction allows quantum non-demolition (QND) measurements of the qubit in the $\sigma_z$ basis through probing of the qubit-state-dependent cavity resonance.  \add{This measurement architecture is routinely used for projective measurements in the qubit basis}, represented by the projection operators $\Pi_{\pm,z}$. \addk{These measurements} achieve measurement fidelities in excess of $95\%$  with the predominant sources of infidelity arising from qubit transitions \cite{slic12, sank16} that occur during the finite duration of the measurement \cite{Evan14,Rist12,John12,Mack15,Wall05,sank16, chen16}.

% {\bf We need to think of a condensed description of this measurement}  These measurements are described by the POVM operators
%\begin{eqnarray}
%\Omega_V = (2 \pi a^2)^{-1/4} e^{[(V-\sigma_z)^2/4a^2}
%\end{eqnarray}

We can make more general projective measurements by combining measurements in the $\sigma_z$ basis with arbitrary rotations ($R_x^\theta,\ R_y^\theta$) about the $x$ and $y$ axes of the qubit. For example, a projective measurement along the axis that forms an angle $\theta$ with the $z$ axis and azimuthal angle $\phi = 0$ can be performed through the \add{following operations}, $\Pi_{ \pm,\theta} = R^{-\theta}_y\ \Pi_{\pm,z}\ R^{\theta}_y$ (Fig.\ \ref{fig:rho}b). In the following these projective measurements will constitute the POVMs, $\Omega_{ \pm,\theta} = \Omega_{ \pm,\theta}^\dagger \equiv \Pi_{ \pm,\theta}$, for which we will test the predictions, Eqs.\ (\ref{eq:cmprob}, \ref{eq:pqsprob}). If the qubit is described by a diagonal density matrix $\rho\addk{(t)}$, the probability of obtaining eigenvalue $+1$ \add{(associated with the state $|0\rangle$)} from such a measurement is given by,
\begin{eqnarray}
P_\rho(+,\theta)=\rho_{00}\addk{(t)} \cos^2\bigg(\frac{\theta}{2}\bigg)
+ \rho_{11}\addk{(t)} \sin^2\bigg(\frac{\theta}{2}\bigg).\label{Ptheta} %P_\rho(\theta)=\rho_{00} \cos(\frac{\theta}{2})^2 + \rho_{11} \sin(\frac{\theta}{2})^2  \label{Ptheta}
\end{eqnarray}

In Figure \ref{fig:rho}c we test the predictions given by Eq.\ (\ref{Ptheta}) for different values of $\rho\addk{(t)}$. To prepare different mixed states, we apply a  qubit rotation pulse $R_y^{\varphi}$  followed by a projective measurement $\Pi_{\pm, z}$. When the result of this measurement is ignored, the projective measurement decoheres the system and prepares the qubit in a diagonal mixed state in the qubit basis eigenstates $|0(1)\rangle \equiv | +(-)z\rangle$ with $\rho_{00}\addk{(t)}$ and $\rho_{11}\addk{(t)}$  determined by the initial rotation angle $\varphi$ and $T_1$ decay during the first measurement. Following this preparation, we make projective measurements at different angles $\theta$  to determine $\tilde{P}(+,\theta) \addk{\equiv} N_+/(N_++N_-)$ from the number of positive (negative) eigenvalue results $N_+$ ($N_-$). The projective measurements $\Pi_{\pm, \theta}$ are subject to infidelities originating predominantly from $T_1$ decay during the $t_\mathrm{m} = 400$ ns projective measurement.  This results in a $\theta$-dependent measurement fidelity that is given by the overlap of the $\Pi_{\pm, \theta}$ eigenstates and the qubit excited state, $\mathcal{F}_\theta = 0.99-\sin^2 (\theta/2) (1-e^{-t_\mathrm{m}/T_1})$ and ranges from $0.945$ when $\theta = \pi$ to $0.99$ when $\theta = 0$. The maximum readout fidelity of $0.99$ arises from residual overlap of the measurement distributions.  After correcting for the measurement fidelity, the predictions given by $\rho\addk{(t)}$ are in good agreement with the measured probabilities as shown in Figure \ref{fig:rho}c.

\begin{figure}\begin{center}
\includegraphics[angle = 0, width =.48\textwidth]{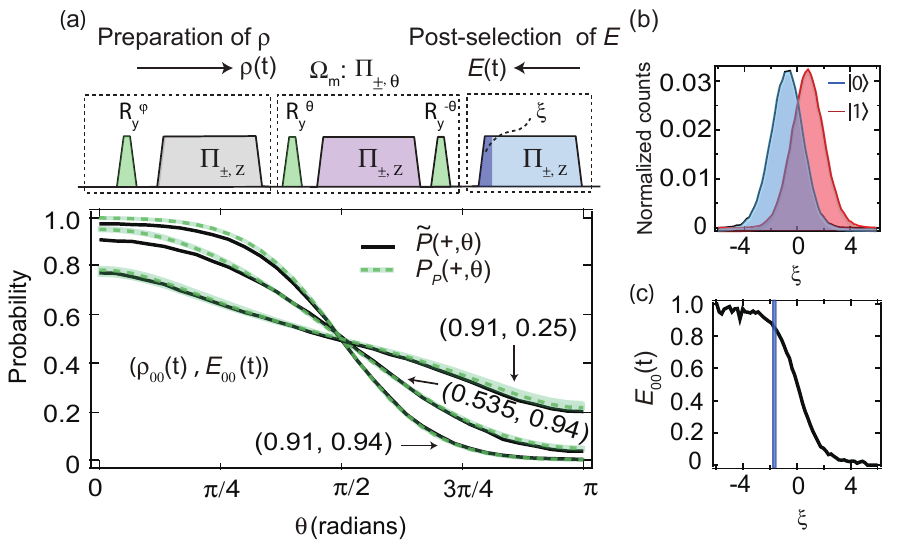}
\end{center}
\caption{ \label{fig2} \add{(a)}  Experimental sequence and comparison of experiments with the predictions of projective qubit measurement outcomes along $\theta$ using both $\rho$ and $E$ for three different mixed states. After the $\Pi_{\pm,\theta}$ measurement, a 30 ns integration of a readout signal is used to determine $E\addk{(t)}$. \add{The solid lines are the measured probability based on over $5\times10^4$ experimental iterations for each value of $\theta$,} and the dashed lines are the theoretical prediction from Eq.\ (\ref{eq:Ppthetared}). (b) Histograms of the integrated 30 ns readout signals $\xi$ for the qubit prepared in the ground (blue) and excited (red) state which are used to create \addk{the map between  $\xi$ and $E_{00}\addk{(t)}$},  shown in panel (c).
% $E_{00}\addk{(t)}$ as a function of the the signal $\xi$ which is calculated from the histograms. 
 \add{The finite width of the post-selection window for determination of $E\addk{(t)}$ \cite{eref}, shown as the blue vertical line, gives rise to a range of values for the theory predictions which are indicated in (a) by the thickness of the faint green curves.}}
\end{figure}

We now address how the subsequent continuous probing of the qubit in the $\sigma_z$ basis, as depicted in Figure \ref{fig2}a, yields our smoothed predictions for the outcomes of the projective measurements $\Pi_{\pm, \theta}$. After the dispersive interaction, the phase of the coherent probe field depends on the qubit state, and the time integral $\xi$ of the measured $Q$-quadrature is Gaussian distributed with opposite mean values for the states $|0(1)\rangle$. In Fig \ref{fig2}b, we show the experimentally obtained distributions $P(\xi|0)$ and $P(\xi|1)$, where we have normalized the integrated signal to have mean values $\pm 1$ for the two qubit states. The Gaussian widths are significant for short probing times and become much narrower when the system is probed for longer.
For a given measured signal $\xi$, we can extract the values $P(\xi|0)$ and $P(\xi|1)$, i.e., the probability of the measured signal conditioned on the state. By Bayes' rule, these are precisely the factors multiplying the prior probabilities $\rho_{nn}\addk{(t)}$ to yield the classical smoothing theory. \textit{I.e}., if we disregard the effect of qubit decay during the probing, they yield the values of $E_{00}\addk{(t)}$ and $E_{11}\addk{(t)}$ in Eq.(2),
\begin{eqnarray}\label{Emap}
E_{00}\addk{(t)}=\frac{P(\xi|0)}{P(\xi|1)+P(\xi|0)}, \
E_{11}\addk{(t)} =\frac{P(\xi|1)}{P(\xi|1)+P(\xi|0)},
\end{eqnarray}
where we have applied a common normalization factor, leading to Tr($E$)$=1$. Fig \ref{fig2}c shows how the inferred normalized value of $E_{00}$\addk{(t)} ($E_{11}=1-E_{00}$) depends on the measured signal $\xi$.  The continuous probing constitutes a QND measurement of the qubit state, and the accumulated back-action on the qubit state populations in the forward propagation of $\rho$ \cite{koro11} amounts to the same factors---which confirms that the evolution of $E$ is, indeed equivalent to the evolution of $\rho$ (the QND back-action is equal to its adjoint).

For a projective measurement in the qubit basis $(\theta = 0)$ at time $t$, $\rho(t)$ leads to the prediction $P_{\rho}(0) = \rho_{00}(t)$, while the pair of matrices $\bigl(\rho(t),E(t)\bigr)$ implies
\begin{equation}
\addk{P_P(0)\equiv P_P(+,0) =} \frac{\rho_{00}(t)E_{00}(t)}{\rho_{00}(t)E_{00}(t)+\rho_{11}(t)E_{11}(t)}.
\end{equation}
If the values of $P_P(0)$ and \addk{$P_P(1)\equiv P_P(-,0)=1-P_P(0)$} could be interpreted as refined populations of a classical mixture of the two qubit states at time $t$, the projective measurement, corresponding to $\Pi_{+,\theta}$ would have the probability
\begin{eqnarray}\label{Pcmtheta}
P_P^{cm}(+,\theta)=P_P(0) \cos^2\left(\frac{\theta}{2}\right)
+ P_P(1) \sin^2\bigg(\frac{\theta}{2}\bigg),
\end{eqnarray}
while insertion of the projection operators $\Pi_{\pm,\theta}$ for $\Omega_m$ in \eqref{eq:pqsprob} yields the expression
\begin{equation}\label{eq:Ppthetared}
P_P(+,\theta) = \frac{P_\rho(+,\theta) P_E(+,\theta)}{P_\rho(+,\theta) P_E(+,\theta) + P_\rho(-,\theta)P_E(-,\theta)},
\end{equation}
where $P_\rho(+,\theta)$ is given in \eqref{Ptheta}, and we have introduced the formally similar $P_E(+,\theta) = E_{00} \cos^2\left(\frac{\theta}{2}\right) + E_{11} \sin^2\left(\frac{\theta}{2}\right)$ and $P_\rho(-,\theta)=1-P_\rho(+,\theta)$, $P_E(-,\theta)=1-P_E(+,\theta)$.

In our experiment, the signal related to $E\addk{(t)}$ is obtained from \addk{additional probing that immediately follows the} measurement $\Pi_{\pm, \theta}$. $E\addk{(t)}$ is given by the Eq.\ ($\ref{Emap}$) and depicted in Figure \ref{fig2}c.   %We sort these signals into bins of width 0.19. E is determined by the averaged signals which were post-selected in a bin based on the equation \ref{Emap}.
In Figure \ref{fig2}a, we display our experimental results that test the prediction of Eq.\ ($\ref{eq:Ppthetared}$) for three different combinations of $\rho\addk{(t)}$ and $E\addk{(t)}$ \cite{eref}.  %To determine $E$, we postselect on the final integrated signal $\xi$ (within a window of width $0.19$ as shown in Figure \ref{fig2}b).
The experimental and theoretical curves show good agreement and highlight how information before and after the projective measurement contribute to the smoothed prediction.

\begin{figure}\begin{center}
\includegraphics[angle = 0, width =.5\textwidth]{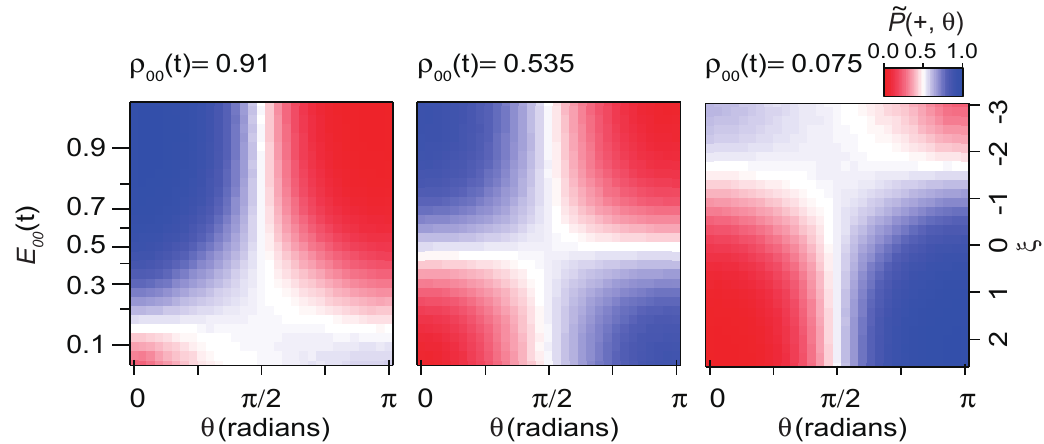}
\end{center}
\caption{ \label{fig3} The experimentally determined $\tilde{P}(+,\theta)$ as function of $\theta$ and $E_{00}\addk{(t)}$ is shown for three different initial states with $\rho_{00}\addk{(t)} = \{ 0.91,\ 0.535, 0.075 \}$ (left to right). The figures show how information accumulated after the projective measurement supplements $\rho$ to further bias or unbias the measurement outcome probabilities.}
\end{figure}

\begin{figure}\begin{center}
\includegraphics[angle = 0, width =.4\textwidth]{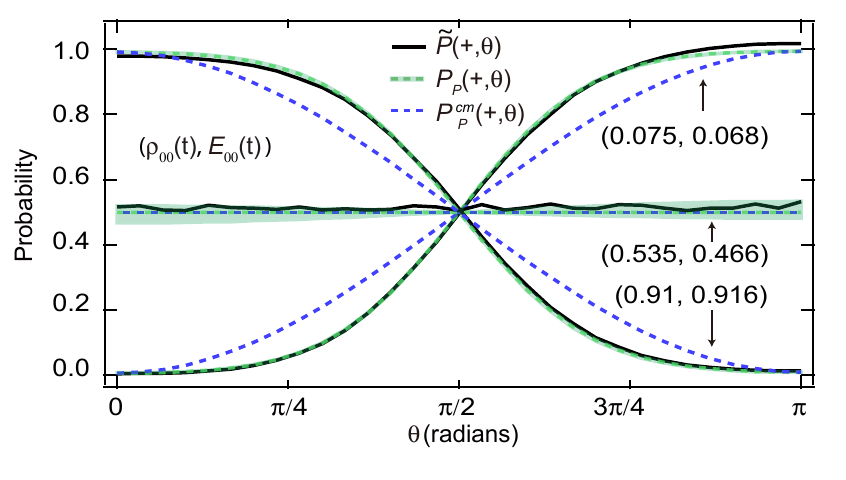}
\end{center}
\caption{ \label{fig4} Comparison of the $\tilde{P}(+,\theta)$ \add{(solid lines, based on over $5\times10^{4}$ experimental iterations with statistical errors of order $4\times 10^{-3}$)} to the smoothed prediction $P_P(+, \theta)$ (green dashed \add{with green bands indicating the theory uncertainty \cite{eref}}) and to the prediction based on a classical mixture with the smoothed state occupations, $P_P^\mathrm{cm}(+, \theta)$ (blue dashed).  We display results for three different values of $(\rho\addk{(t)}, E\addk{(t)})$; $\{ (\rho_{00}\addk{(t)}, E_{00}\addk{(t)}) =  (0.91,0.916),\ (0.535,0.466),\ (0.075,0.068)\}$.}
\end{figure}

Figure $\ref{fig3}$ summarizes our experimental results, showing the measured  $\tilde{P}(+,\theta)$ as a function of the angle $\theta$ and the post-selected value of $E_{00}\addk{(t)}$ (the corresponding values of the integrated signal $\xi$ are given on the right hand axis in the figure). Results are shown for three different density matrices $\rho\addk{(t)}$ prior to the projective measurement along the direction $\theta$. For $\theta=\pi/2$ both conventional and smoothed predictions assign unbiased probabilities $0.5$ to the outcomes $\pm,\theta$. For any $\theta$ and for all three values of $\rho\addk{(t)}$, a certain value of the probing signal after the projective measurements results in an unbiased smoothed prediction  $P_P(+,\theta)=0.5$. This amounts to an increased uncertainty about the outcome and it happens because the subsequent probing of the system is at loggerheads with the prior state $\rho\addk{(t)}$ (e.g., $\rho_{00}\addk{(t)} = 0.91$, $E_{00}\addk{(t)} = 0.25$, cf., Fig.\ 2c). Conversely, when the $\rho\addk{(t)}$ and $E\addk{(t)}$ are similar (e.g., $\rho_{00}\addk{(t)} = 0.91$, $E_{00}\addk{(t)} = 0.94$, cf., Fig.\ 2c), the later probing "confirms" the prediction by $\rho\addk{(t)}$, and thus enhances the probability of the most likely outcome of the projective measurement. These trends are most clearly observed in Figure \ref{fig4}, where we compare the measurement probabilities $\tilde{P}(+,\theta)$ to the smoothed prediction $P_P(+,\theta)$ and the classical mixture interpretation $P_P^\mathrm{cm} (+,\theta)$. Notably, the figure shows a clear disagreement of the experimental data with the classical mixture interpretation.

% \section{Conclusion}
In conclusion, we have presented a description of a quantum system, evolving without developing coherences, and hence, both prior and posterior information about the  system are represented by diagonal matrices. While the theory of smoothing yields probabilities in better agreement with predictions for the outcome of measurements in the eigenstate basis, we have shown that these probabilities do not permit a classical mixture interpretation of the (past) quantum state.

\add{While our experimental observations are at variance with a classical mixture interpretation, they have a physical explanation: The intervening projective measurements explicitly break the notion of classical mixtures, because their back-action leaves the system in states with non-vanishing coherences in the $|\pm,z\rangle$ basis.  Notably,  however, the matrices $\rho(t)$ and $E(t)$ do not refer to the specific measurement at time $t$, and our central question was if the state, known to us from the data leading to $\rho(t)$ and $E(t)$ is equivalent to a classical mixture. It is not.}

At a more foundational level, our work dismisses simple "hidden variable theories" that equate eigenstates of incoherent ensembles with hidden "true" states of the system, and it offers an illustration of the problematic character of macrorealism \cite{legg85} which separates the evolution of quantum states and the measurements performed. Rather than demonstrating an explicit statistical violation of the Bell \cite{Bell65}, CSCH \cite{CSCH81}, or Leggett-Garg \cite{Groe13,pala10,will08,gogg11,whit15} inequalities, we have merely shown the failure of the simplest preconceived probabilistic classical mixture interpretation of the quantum description, and we have shown that the pair of matrices $\rho(t)$ and $E(t)$ offers a satisfactory account of the outcomes of past measurements on a quantum system.

We acknowledge P. Harrington and N. Foroozani for discussions and assistance with the manuscript and G. Zhao, L. Xu, and L. Yang for fabrication assistance. This research was supported in part by the John Templeton Foundation and the Sloan Foundation and used facilities at the Institute of Materials Science and Engineering at Washington University.  K.M. acknowledges support from the Villum Foundation.

Correspondence and requests for materials should be addressed to K.W.M. (murch@physics.wustl.edu)


\begin{thebibliography}{38}
\bibitem[{The experimental set-up is similar to previous work
  \cite{tan15}()}]{EXP}
The experimental set-up is similar to previous work \cite{tan15},
  \bibinfo{note}{with qubit transition frequency
  $\omega_\mathrm{q}/2\pi=5.44888$ GHz and coherence properties, $T_{1}=8$
  $\mu$s, $T_{2}^{\ast}=9.5$ $\mu$s. The qubit is dispersively coupled to a
  wave-guide cavity with resonance frequency $\omega_\mathrm{c}/2\pi =6.76208$
  GHz and linewidth $\kappa/2\pi = 6.3$ MHz and dispersive coupling
  $\chi/2\pi =-0.8$ MHz.}
\bibitem[{ere()}]{eref}
\bibinfo{note}{To study specific values of $E$ we post-select on the final
  measurement $\xi$ within a window of width $0.19$. \addk{The width of this window results in a range of theory curves which are depicted as a finite width band.}}.
\end{thebibliography}
\end {document}